\documentclass{article}
\usepackage{cite}
\usepackage{amsmath,amssymb,amsfonts}
\usepackage{algorithmic}
\usepackage{graphicx}
\usepackage{textcomp}
\usepackage{xcolor}
\usepackage{color}
\usepackage[utf8]{inputenc}
\usepackage{algorithm}

\def\BibTeX{{\rm B\kern-.05em{\sc i\kern-.025em b}\kern-.08em
    T\kern-.1667em\lower.7ex\hbox{E}\kern-.125emX}}

\newcommand{\code}[1]{{\ttfamily #1}}
\newcommand{\power}{POWER9 cluster}

\begin{document}

\providecommand{\keywords}[1]
{
     \small 
       \textbf{\textit{Keywords---}} #1
}

\title{Parallel SFC-based mesh partitioning and load balancing\footnote{
© 2020 IEEE. Personal use of this material is permitted. Permission from IEEE must be obtained for all other uses, in any current or future media, including reprinting/republishing this material for advertising or promotional purposes, creating new collective works, for resale or redistribution to servers or lists, or reuse of any copyrighted component of this work in other works.
DOI:10.1109/ScalA49573.2019.00014
}}
\author{ Ricard Borrell, Guillermo Oyarzun, Damien Dosimont, \\ 
                and Guillaume Houzeaux \\
        {\textit{Barcelona Supercomputing Center}}\\
       {Barcelona, Spain}
}

\maketitle

\begin{abstract}
Modern supercomputers allow the simulation of  complex phenomena with increased accuracy. Eventually, this
requires finer geometric discretizations with larger numbers of mesh elements. In this context, and extrapolating 
to the Exascale paradigm, meshing operations such as generation, adaptation or partition, become a critical  
within the simulation workflow. In this paper, we focus on mesh partitioning. In particular, we present some improvements 
carried out on an in-house parallel mesh partitioner based on the Hilbert Space-Filling Curve.

Additionally, taking advantage of its performance, we present the application of the SFC-based partitioning for dynamic
load balancing. This method  is based on the direct monitoring of the imbalance at runtime and the subsequent re-partitioning
of the mesh. The target weights for the optimized partitions are evaluated using a least-squares approximation considering all
measurements from previous iterations. In this way, the final partition corresponds to the average performance of the computing 
devices engaged. 

\end{abstract}


\section{Introduction}

Mesh partitioning for the parallelization of PDE solvers is traditionally formulated as a graph partition problem, which derives in
a well-studied NP-hard problem generally addressed using multilevel heuristics composed of three phases: coarsening, partitioning, and
un-coarsening. Publicly available libraries such as ParMetis~\cite{Metis} or PT-Scotch~\cite{PTscotch} implement different 
variants of them. An alternative approach are the geometric partitioning techniques, which obviate topological interactions 
between mesh elements and perform the partition considering only its spatial location. A Space-Filling Curve (SFC) is 
generally used to map the mesh elements into a 1D space, which is then split it into equally weighted sub-segments. 
There are different definitions of SFC, such as the Hilbert or Peano curves \cite{Sagan1994}. 
The common goal is to preserve locality, i.e., preserve the closeness among points on their 1D projections and vice versa. 
A significant advantage of  geometric approaches is that they can be fast and easy to parallelize, 
especially when compared to graph partitioning methods. However, while a well balanced partition can be easily achieved if the SFC 
granularity is fine enough, the interface between subdomains - measured in terms of edge-cuts in the graph methods- is not explicitly 
minimized but it's only treated indirectly through the SFC locality property. 

\begin{figure*}[ht!]
   \centering
     \includegraphics[width=0.8\textwidth]{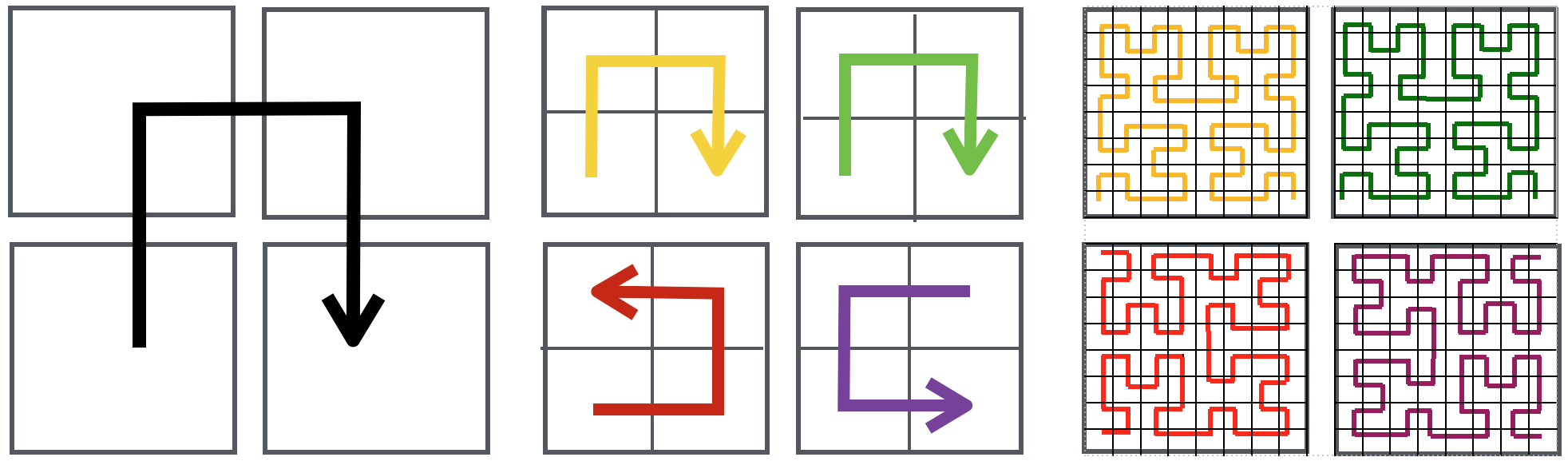}
    \caption{Hilbert SFC parallel generation. a) Initial coarse grid definition, b) orientation of each parallel process to continue the recursion, c) recursive generation of SFC in each parallel process.}
     \label{sfcpargen}
 \end{figure*}
 
In this paper we present some improvements carried out on the SFC-based algorithm that we previously presented in~\cite{BOR18}. 
On that work, we showed results for parallel mesh partitions using up to $~4$K CPU-cores on the Blue Waters supercomputer from the 
University of Illinois. We asserted as well that the partition is independent (discounting round-off errors) of the number of 
parallel processes used to compute it. In the new version of the algorithm, we have eliminated a restriction imposed on the number of
MPI processes used to carry out the partition. In particular, to generate a coherent Hilbert SFC in parallel we required that the number 
of processes was a power of $2^d$, being $d$ the dimension of the domain space. This restriction has been overcome  in the current 
version of the algorithm. 

Additionally, in this paper we  present our approach to use the SFC-partition for load balancing purposes. This balancing 
method is based on the direct measurement of the imbalance at runtime and the subsequent re-partitioning of the mesh. The target
weights for the optimized partitions are evaluated using a least-squares approximation considering all  measurements from previous
iterations. In this way the final partition corresponds to the average performance of each computing device engaged. 

The SFC based partitioner developed in this paper and the physics solvers used in the test cases are both integrated into
Alya~\cite{Houzeaux09,Houzeaux11}: the high-performance computational mechanics code developed at the Barcelona Supercomputing Center. 
The physics solvable with the Alya system include incompressible/compressible flows, solid mechanics, chemistry, particle transport, 
heat transfer, turbulence modeling, electrical propagation, etc. Alya aims at massively parallel supercomputers~\cite{vazquez2016};
its parallelization includes both the MPI and OpenMP frameworks, as well as heterogeneous options with accelerators. 
Alya is one of the twelve simulation codes of the Unified European Applications Benchmark Suite (UEABS) of PRACE~\cite{ueabs} and thus complies 
with the highest standards in HPC.

The structure of the paper is as follows: Section~\ref{sec:sfc} contains a short overview of the Hilbert SFC; in 
Section~\ref{sec:sfcpart} we summarize our in-house algorithm for parallel partitioning based on SFC. Section~\ref{sec:optimization} 
contains the improvements proposed. Section~\ref{sec:loadbalancing} the implementation of the load-balancing strategy. 
Finally, general conclusions are outlined section~\ref{sec:conclusions}.

\section{Hilbert Space Filling Curve}
\label{sec:sfc}

Space-Filling Curves (SFC) are used to map a multi-dimensional space into a one-dimensional
space preserving locality. There are many possible definitions of SFC based on different mapping options, among 
them the well-known Peano~\cite{Peano1990} and Hilbert~\cite{Hilbert1970} definitions. Sagan~\emph{et al.}~\cite{Sagan1994} 
give a complete overview of  different options. In this paper, the Hilbert SFC is selected because of its good locality 
preservation, however, switching to another curve definition would be straightforward in our implementation.

The Hilbert curve can be generated as a geometric recursion, Figure~\ref{sfcpargen} represents the three initial steps for the
2D case. It also illustrates the parallel multilevel strategy used to generate the curve: an initial coarse grid is defined 
and its bins a ordered according to the SFC; then each  parallel process continues the recursion 
within a \textit{coarse bin}. Following the proper orientation within each coarse bin, the resulting SFC generated in parallel is 
the same that would be obtained sequentially. The parallel generation of an SFC has two obvious advantages: it is faster than the 
sequential version and, since there is more memory available, a higher level of refinement can be achieved. This second aspect 
eventually determines the granularity of the balancing process. Recalling that the ordering of the coarse bins (associated 
to parallel processes) is based on the Hilbert SFC, this ordering restricts the number of parallel processes, $P$, that should be a 
power of $2^d$, where $d\in\{2,3\}$ is the dimension of the domain. As explained in section~\ref{sec:optimization}, we have overcome 
such restriction by \textit{over-decomposing} the problem.

 \section{Overview of the SFC based mesh partitioning algorithm}
\label{sec:sfcpart}

\begin{figure}[ht!]
  \centering
    \includegraphics[width=0.45\textwidth]{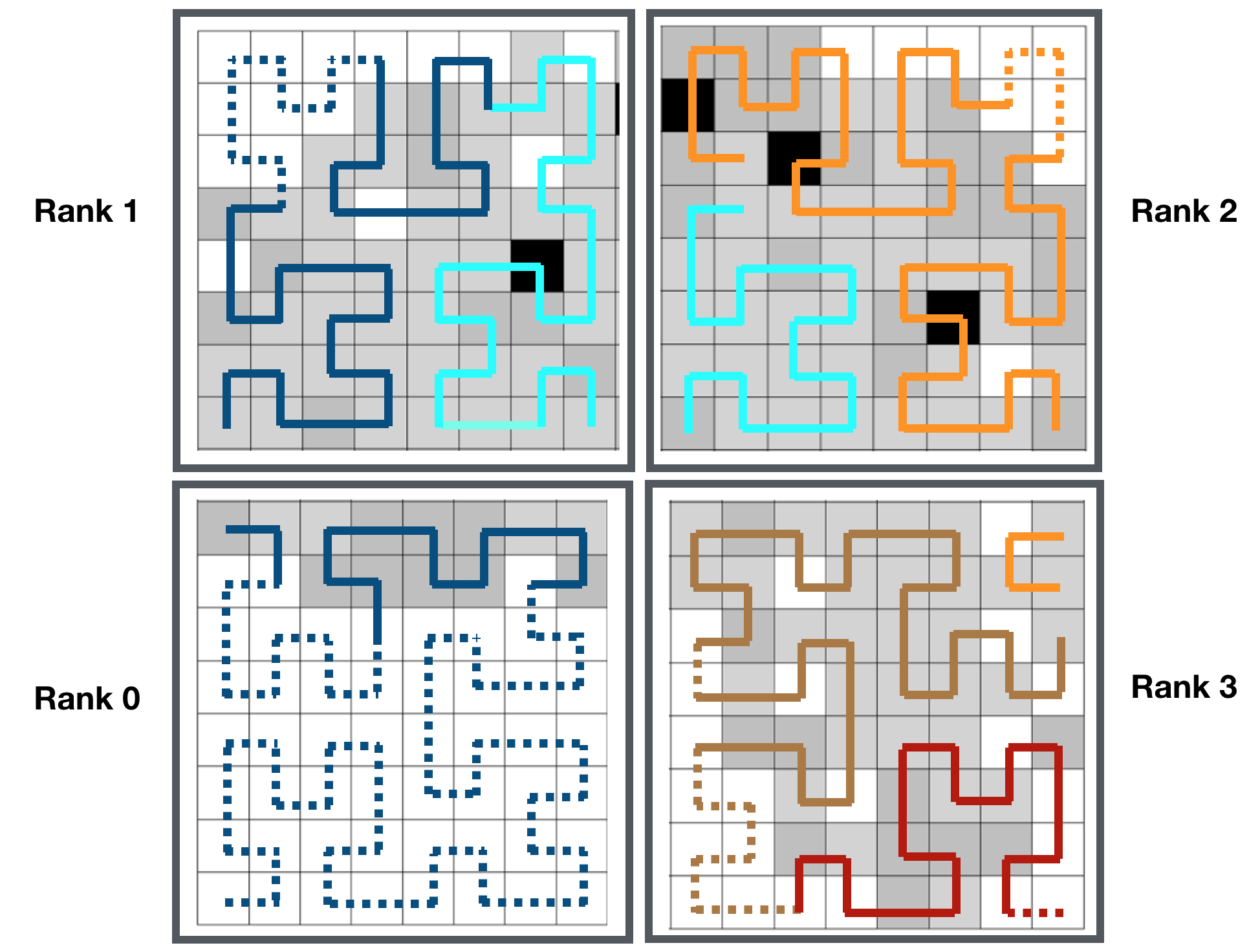}
   \caption{Parallel mesh partitioning using SFC.}
    \label{fig:sfcpar}
\end{figure}

 The steps of the SFC-based mesh partitioner are the following: i) a bounding box is defined enclosing the domain; ii) a regular grid 
 is defined inside the bounding box; iii) the bins of the regular grid are weighted according to the elements contained in them; iv) 
 the SFC curve is used to project the bins to a 1D space; and v) the 1D partitioning problem is solved. In the parallel implementation, 
 we divide the bounding box into sub-boxes, and each parallel process performs a local partition within
 its sub-box. Connecting the local partitions, we obtain a coherent partition of the overall mesh. This parallelization strategy is 
 illustrated in Figure~\ref{fig:sfcpar}. In this  example, the mesh is divided into five parts employing four parallel processes.   
 
The steps of the parallel partitioning algorithm are outlined in Algorithm~\ref{alg:parsfc}. See ~\cite{BOR18} for more details.

As initial condition of the parallel partitioning algorithm, we assume that the mesh cells are distributed among $P$ parallel processes. 
Regarding the data transfers, there are three collective communications. In step $1$ a \code{MPI\_Allreduce} communication is used to
evaluate the limits of the bounding box enclosing the mesh. In step $2$ a \code{MPI\_Alltoall} is used to set up the communication 
requirements for the bins weight redistribution and, finally, in step $3$ a \code{MPI\_Allgather} is used to obtain the weight 
distribution across the coarse bins on each parallel process. Moreover, two point-to-point communications are necessary: to redistribute 
the weight of the bins (step $2$), and to obtain the result on each parallel process according to the elements it initially holds (step $5$).

\begin{algorithm}[h!]
\caption{Parallel SFC-based partitioning} 
\label{alg:parsfc}
\begin{algorithmic}[1]
   \STATE Definition of the bounding box and the cartesian grids to be used
   \STATE Evaluation and redistribution of fine-grid bins weight
   \STATE Gather the coarse-grid weights distribution on all processes   
   \STATE Local partition based on SFC
   \STATE Redistribution of the result of the local-bins partition
   \STATE Infer mesh partition
\end{algorithmic}
\end{algorithm}	

 In our previous work~\cite{BOR18}, we presented a scalability study of the algorithm on the AMD based Blue Waters supercomputer 
 from the University of Illinois. We succeeded in partitioning a mesh of $30$M elements in approximately one second by using $8$ CPU-cores,
 and a few cents of a second when running on $4$K CPU-cores. Apart from round-off errors, the sequential and parallel executions of the 
 algorithm produce the same partition if the fine grid resulting from joining the local grids is the same than the one used for the
 sequential partition. However, since a larger grid can result in a better balanced partition with the parallel execution, we can 
 produce more accurate results. We observe an opposite behavior with graph partitioners, where sequential and parallels results are generally
 different, and the parallelization can worsen the result~\cite{Karypis98}.

\section{Extension of the algorithm}
\label{sec:optimization}

\begin{figure}[ht!]
  \centering
    \includegraphics[width=0.45\textwidth]{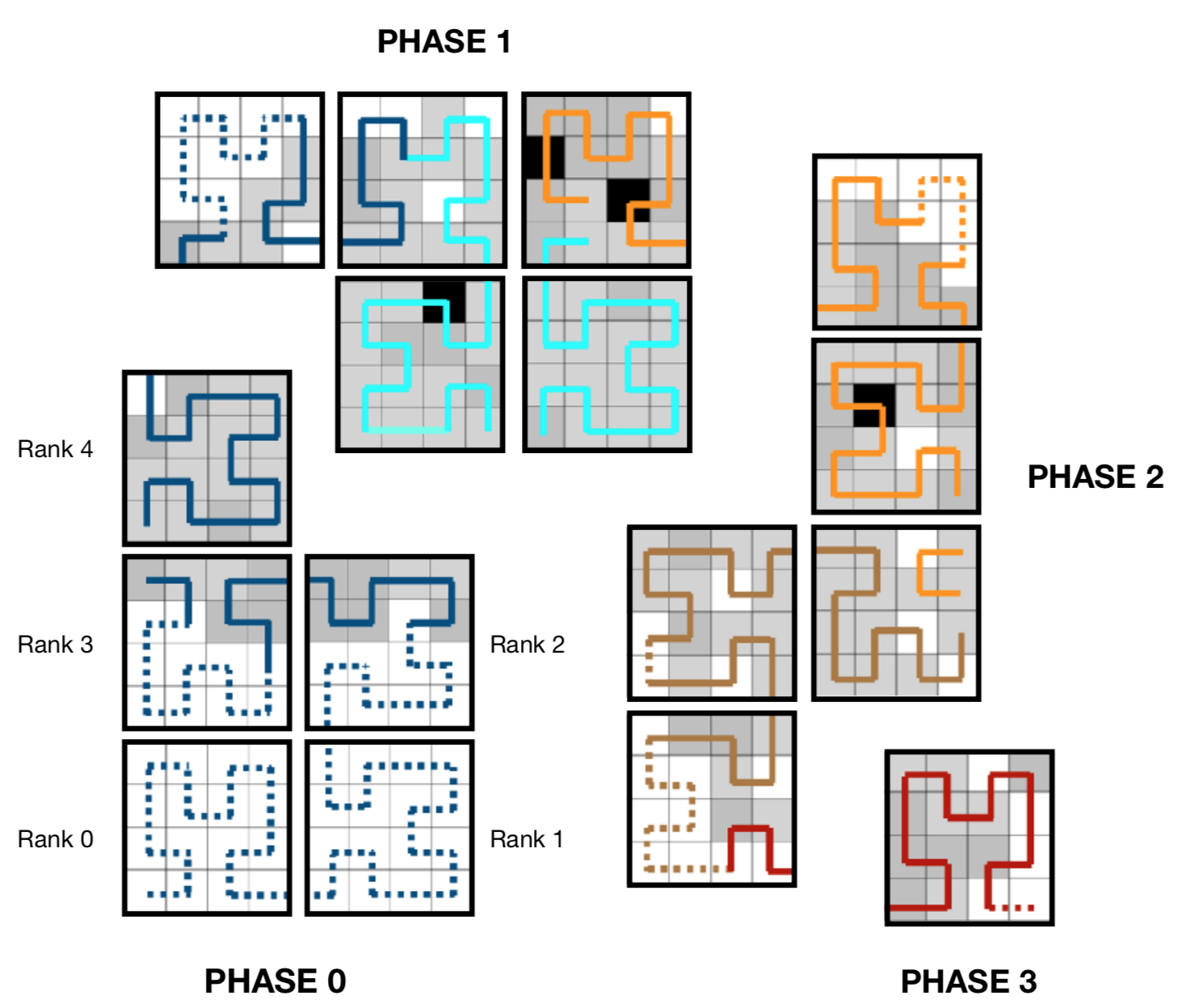}
   \caption{Parallel mesh partitioning using SFC.}
    \label{fig:sfcpar2}
\end{figure}

In the current version of the algorithm, we have eliminated the restriction imposed on the number of parallel processes usable for
the partition. This needed to be $2^{dq}$, being $d$ the dimension of the  domain ($2$ or $3$), and $q$ any natural number. This 
restriction allowed to directly use the SFC-index to assign each coarse bin to a parallel process to generate a coherent SFC 
in parallel. In the current version of the algorithm, we keep such restriction to generate the coarse grid, but we do not impose
that the number of coarse bins and parallel processes is the same. Given the coarse bin with SFC-index $k$,  it 
will be assigned to the process with rank $p=mod(k,P)$. This approach is illustrated in Figure~\ref{fig:sfcpar2}.

The corresponding algorithm is Algorithm~\ref{alg:parsfc2}. In comparison with Algorithm~\ref{alg:parsfc}, this new version requires 
$\frac{N_c}{P}$ \textit{phases} to complete the partition, where $N_c$ is the number of coarse bins. Note that, as with the previous version 
of the algorithm, discounting round-off errors, the final result is independent of the number of parallel processes used to perform the 
partition.

  \begin{algorithm}[h!]
\caption{Parallel SFC-based partitioning} 
\label{alg:parsfc2}
\begin{algorithmic}[1]
   \STATE Definition of a bounding box and cartesian grids to be used
   \STATE Evaluation of the total weight
   \FOR{$1$ \TO \textit{number of phases}}
   \STATE Evaluation and redistribution of fine-grid bins weight
   \STATE Gather the coarse-grid weights distribution on all processes   
   \STATE Local partition based on SFC
   \STATE Redistribution of the result of the local-bins partition
   \STATE Infer mesh partition
   \ENDFOR
\end{algorithmic}
\end{algorithm}	

\begin{figure*}[h]
    \includegraphics[width=0.47\textwidth]{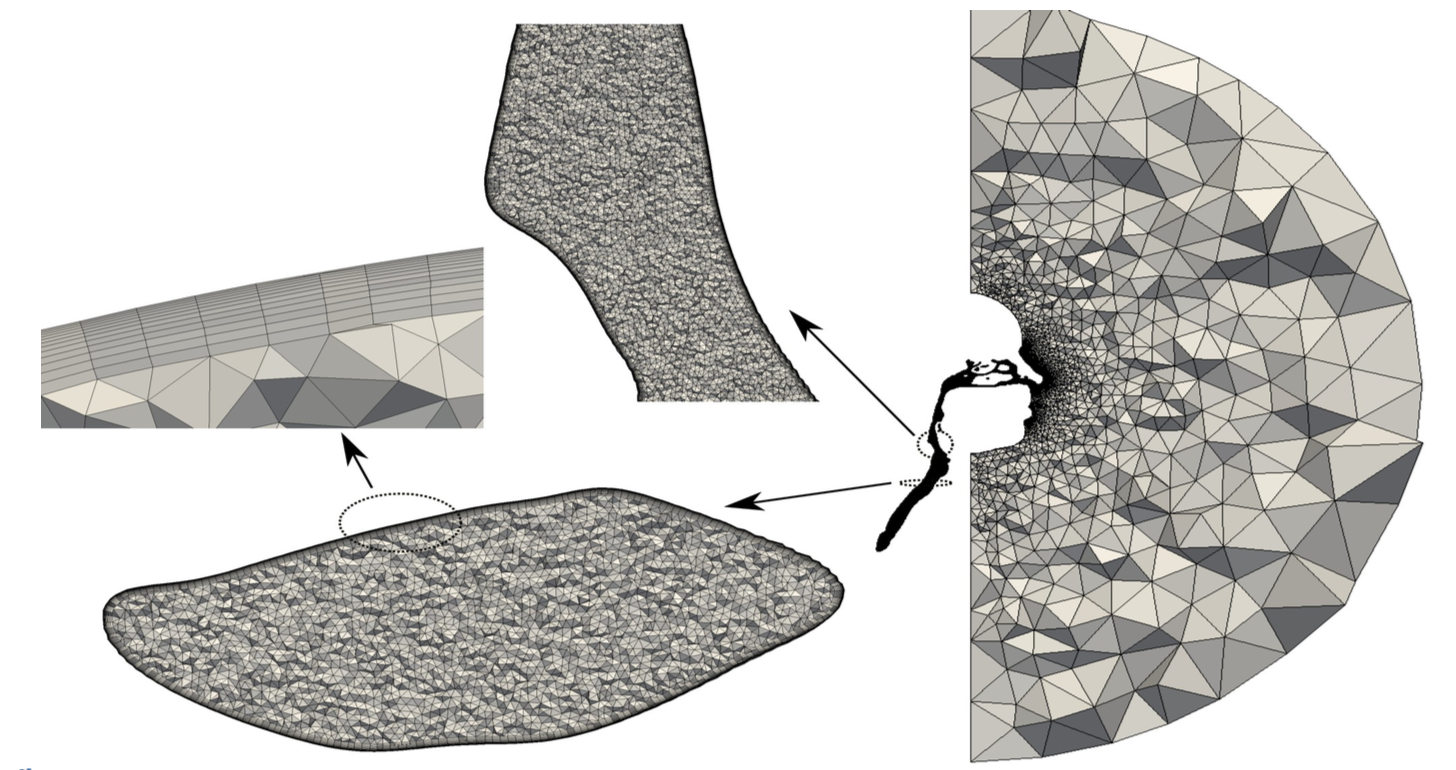}
    \includegraphics[width=0.40\textwidth]{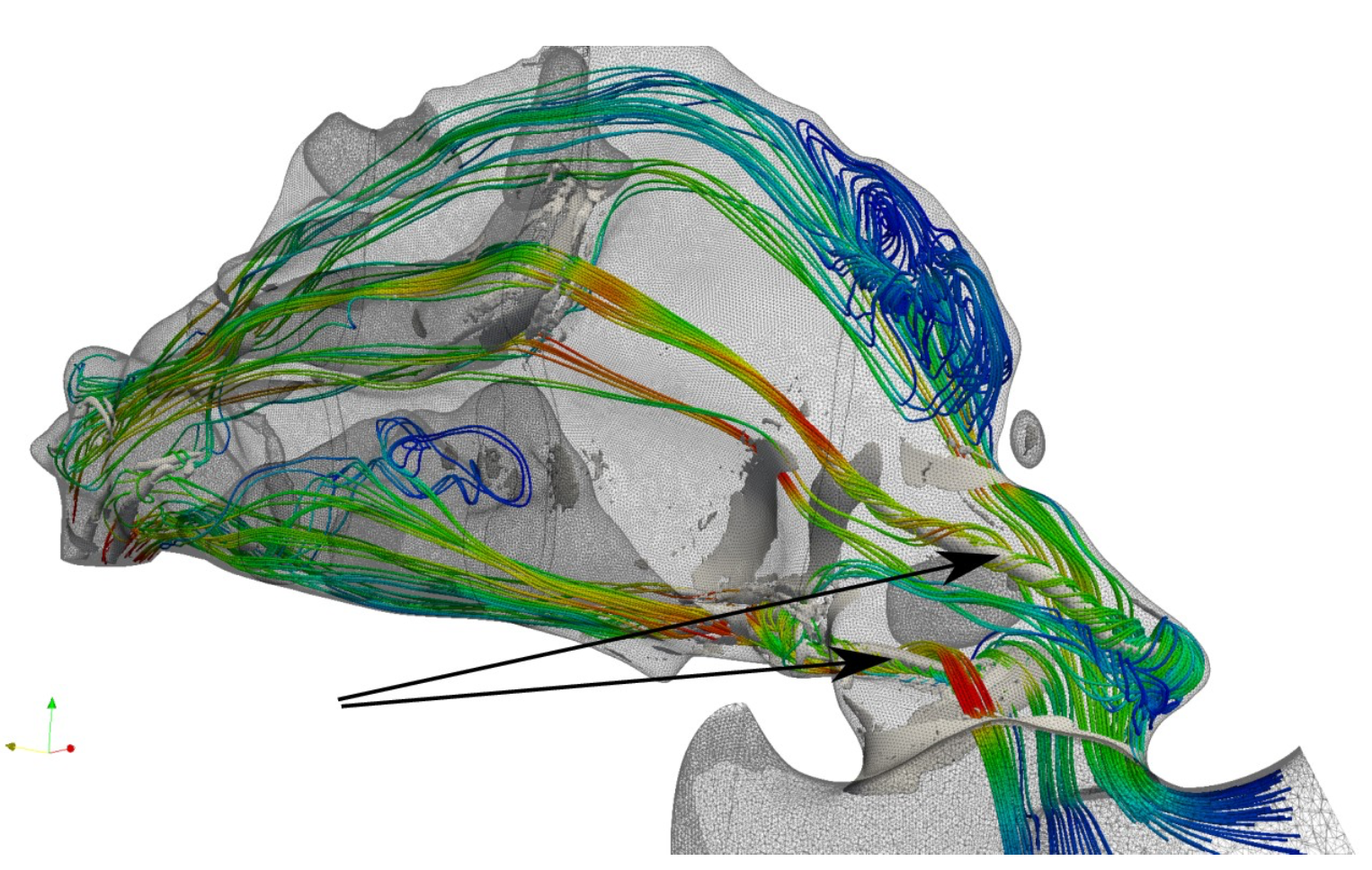}
   \caption{Geometric discretization of the respiratory system (left). Streamlines colored by velocity magnitude 
   and iso-surface of Q-criterion pointing by arrows in the nasal cavities and throat (right).}
\label{fig:sniff}
\end{figure*}

We have performed some preliminary tests in the MareNostrum IV supercomputer from the Barcelona Supercomputing 
Center (BSC). MareNostrum nodes are composed of two Intel Xeo Platinium CPUs, with 24 cores each, 
connected through a Intel Omni-Path network. For the experiments we have considered 
a unstructured mesh of the respiratory system of $~25M$ elements, an illustration of it is shown in Figure~\ref{fig:sniff}, results of the
 corresponding numerical simulation can be found in~\cite{Calmet2016}.

In Table~\ref{tab:opt2} we show results using two different underlying SFCs, defined over a grid of 
$512^3$ and $1024^3$, respectively. On each case we have considered a coarse grid with at least as many bins 
as number of  parallel processes used to perform de partition. Both the CPU time required to perform the partition and the 
respective parallel efficiency are presented. As expected, regarding the scalability, results are better for the larger case. 
Nonetheless, there are still several aspects to be analyzed in detail, such as the optimal granularity of the coarse grid.

\begin{table}
\begin{center}
\begin{tabular}{ | c | c | c | c |}
     \hline
\#CPUs    & $512^3$  & $1024^3$    \\ \hline \hline
     48   & 1,41            & 11,3     \\ \hline
     96   & 0,62 (114\%)       & 4,76 (118\%)    \\ \hline
     192  & 0,43 (82\%)       & 3,73 (76\%)    \\ \hline
     384  & 0,32 (55\%)       & 2.61 (54\%)    \\ \hline
     768  & 0,49 (18\%)       & 1,08 (65\%)    \\ \hline
\end{tabular}
  \caption{Partitioning time (sec.) and parallel efficiency, for two different
  underlying SFCs and different number parallel processes.
  Respiratory system mesh ($25$M).}
\label{tab:opt2}
\end{center}
\end{table}

\section{Load balancing}
\label{sec:loadbalancing}

Heterogeneity, both in the mesh elements, in the algorithms or the hardware, complicates the generation of partitions 
with a uniform work-load distribution. 
Imbalance generates waste of resources since the processes that arrive earlier at the synchronization points remain idle.
The negative effect of imbalance may not show up on scalability tests if the reference execution 
is already imbalanced. However, in many cases the waste of computing resources caused by it is higher than those coming from  
communications overhead.

The a priori estimation of elements weight based on some model is a challenging approach.
Indeed, apart from factors such as the element type or the device where the code runs, there are also indirect factors that 
affect the execution time. For example, the layout in memory of an element with respect to its neighboring elements determines 
the cache misses and thus affects its computing cost. We could instrument the code to measure accurate weights, but it's hardly
possible on an element basis since it would produce a large overhead. In this paper, we present a dynamic load balancing strategy 
based on monitoring computing times in a subdomain basis and correcting the partition accordingly. 

We take advantage of the low latency of the SFC partitioning process to produce corrected partitions according to time measurements. 
We carry out this process iteratively. We specify the correction of the partition through a new input argument added to the SFC partitioner: 
the array $\Lambda[...]$ of dimension $P$. 
When this argument is provided, the target weight for the $i$'th subdomain will be 

\[
\Gamma_i=\lambda_i \frac{W}{P}
\]

\noindent where $\lambda_i$ is the $i$'th component of the array $\Lambda$ and $W=\sum_{i=1}^N{w_i}$ is the sum of the elements 
weight. 
Hence, the mechanism to improve the load 
balancing is not by better adjusting the elements weight but by modifying the target weight per subdomain. 
An obvious requirement is that $\sum_{i=1}^N{\lambda_i} = 1$, so that the weight distributed among the processes is exactly $W$. 

Next, we explain the iterative process to evaluate the correction coefficients $\lambda_i$ that amend the imbalance. First we introduce 
some notation: $t^k_i$ refers to the elapsed time spent by process $i$, in the part of the code under consideration, 
for the $k$'th iteration of the balancing algorithm. The superindex $k$ will be used hereafter to refer to the $k$'th iteration 
of the balancing algorithm. $\bar{t}^k=\frac{\sum_{j=i}^{P}{t_i^k}}{P}$ is the average time.  
$T^k_i=\sum_{j=1}^{i}{t^k_j}$ is the sum of the elapsed time of the first $i$ parallel processes. 
$\Upsilon^k_i=\frac{T^k_i}{\bar{t}^k}$ is the normalized sum. And  $\Lambda^k_i=\sum_{j=1}^{i}{\lambda^k_j}$ 
the sum of the first $i$ correction coefficients. 


The strategy of the algorithm is to evaluate each splitting point independently. Where by splitting point we refer to  
 subdomains division. For each $i$ we find 
the value $\Lambda^*_i$ such that  $\Upsilon^*_i=i$. So we define the splitting point between the first $i$ subdomains 
and the rest. The balancing process consists in evaluating a simple linear regression (SLR) from the existing 
observations $(\Lambda^k_i,\Upsilon^k_i)$ and intersect the regression line with the line $y=i$ to find $\Lambda^{k+1}_i$. Therefore, if 
 the result of the linear regression is $y=\alpha+\beta x$ then
 
 \[
  \Lambda^{k+1}_i = \frac{i-\alpha}{\beta}.
 \]

 \noindent Finally, the new correction coefficients will be 
 
 \[
 \lambda^{k+1}_i=\Lambda^{k+1}_i-\Lambda^{k+1}_{i-1},
 \]
 
\noindent where $\Lambda^{k+1}_0=0$ and $\Lambda^{k+1}_P=P$.

As mentioned earlier, with this method each splitting point is evaluated independently. 
For executions in homogeneous systems, we have observed that $\bar{t}^k$ is practically constant along the iterations (see next subsection); 
i.e., the average time across the parallel processes does not depend on the partition. 
Likewise, $\Upsilon_i^k$ is also independent of the definition of the splitting points for the first $i-1$ subdomains. 
Therefore, the convergence of the sequences $(\Lambda^k_i,\Upsilon^k_i)$ is mutually independent. 
Another property of the method is its resilience to outlier measurements since those are smoothed by the linear regression. 
On the other hand, to accelerate the convergence, we use a weighted linear regression (WLR) instead of an SLR, 
the weight is increased by $50\%$ in each iteration to foster the latest observations.

\subsection{Balancing experiments}

The following experiments have been carried out on the \power{} installed at BSC. 
The cluster consists of 2 login nodes and 52 compute nodes POWER9 AC922 with 2 sockets each, 
512 GB of main memory and 4 Volta V100 compute accelerators. All the 52 nodes are connected via a
Mellanox EDR interconnect fabric. In the following tests only the CPUs have been used.

\begin{figure}[h!tbp]
  \centering
  \includegraphics[width=0.4\textwidth]{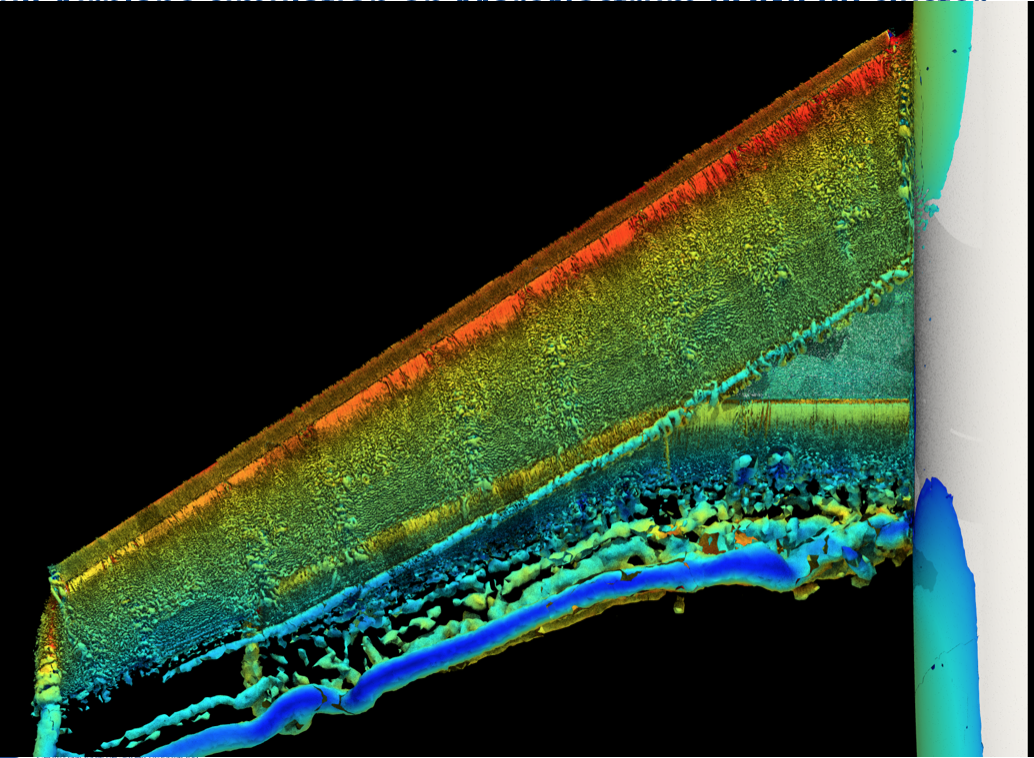}
  \caption{Snapshots of the Q-criterion for the airplane simulation}
  \label{fig:qcriterion} 
\end{figure}

For the load balancing experiments we consider a hybrid mesh consisting of $176$M elements (with tetras, prisms, and pyramids) generated 
for the simulation of the JAXA Standard Model (JSM) of the full airplane.
It has approximately three elements in the inner region of the boundary layer and about $10$ elements in the out layer region. 
Similar mesh resolutions have been used by the authors to simulate comparable flows with success~\cite{rodriguez}. 
 Figure~\ref{fig:qcriterion} depicts instantaneous snapshots of the Q-criterion for the flow around the airplane.

\begin{figure}[!h]
  \centering
    \includegraphics[width=0.40\textwidth]{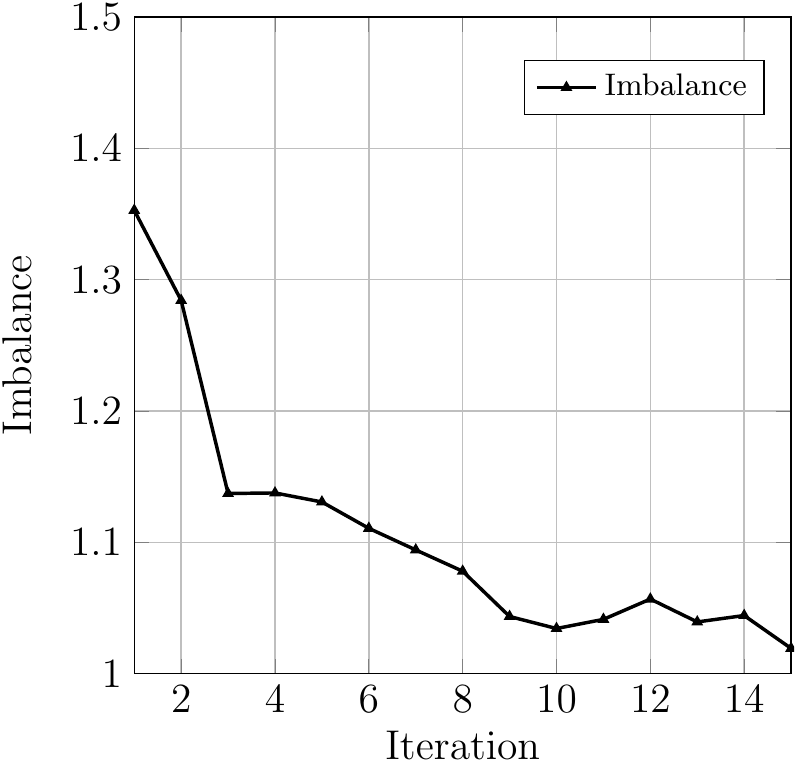}
\caption{Convergence of the balancing process (176M mesh)}. 
\label{fig:conv}	
\end{figure}

The balancing process has been applied to the assembly phase of the time step. For the pure CPU execution this kernel represents 
68$\%$ of the time-step when using 12 nodes (480 CPU-cores), we have considered a hybrid configutation with 240 MPI-processes and two
OmpSs threads associated with each MPI process. The imbalance of an execution is evaluated as 

\[
 I=\frac{max(t_k)}{\bar{t}},
 \]
 
\noindent where $t_k$ is the elapsed time  for process $k$.  
Figure~\ref{fig:conv} shows the evolution of the imbalance through the balancing process for the execution using $12$ nodes. 
The final imbalance avhieved is $0.8\%$. 

\subsection{System instability}
\label{sec:resilience}

\begin{figure*}[!h]
  \centering
    \includegraphics[width=0.6\textwidth]{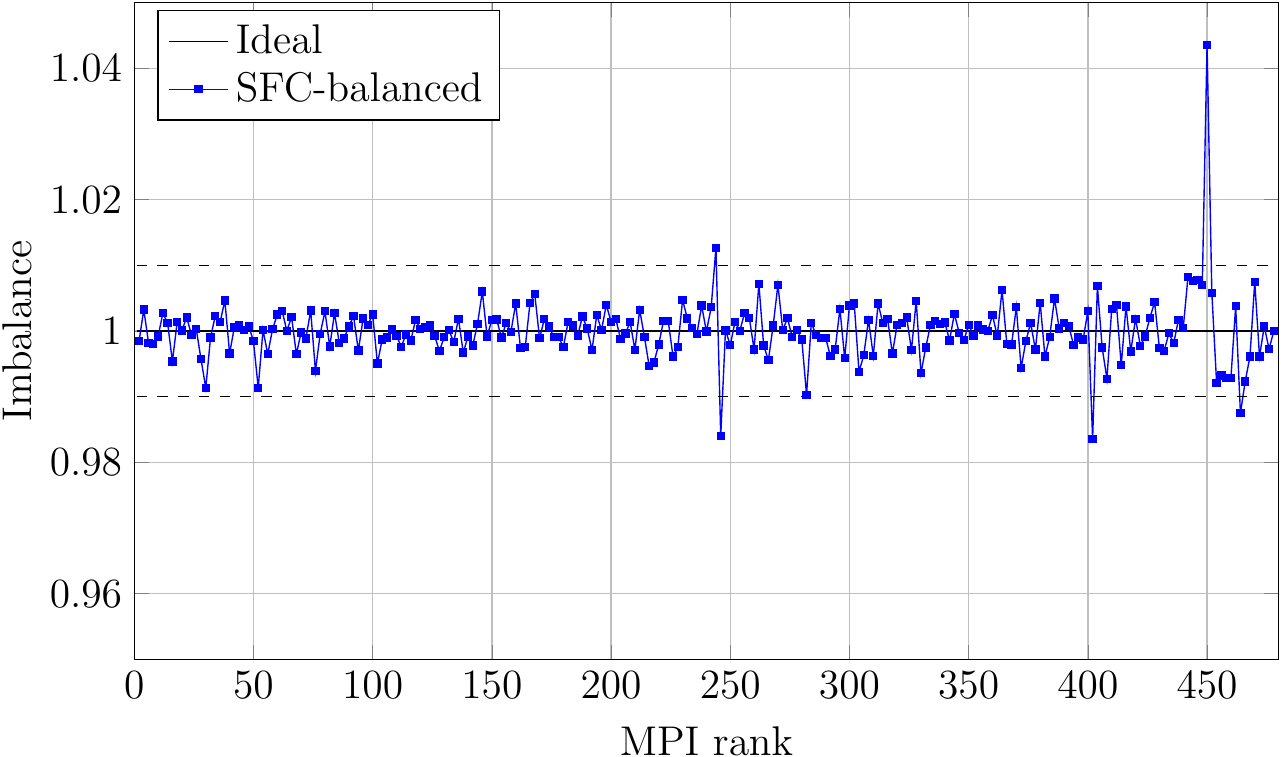}
    \includegraphics[width=0.3\textwidth]{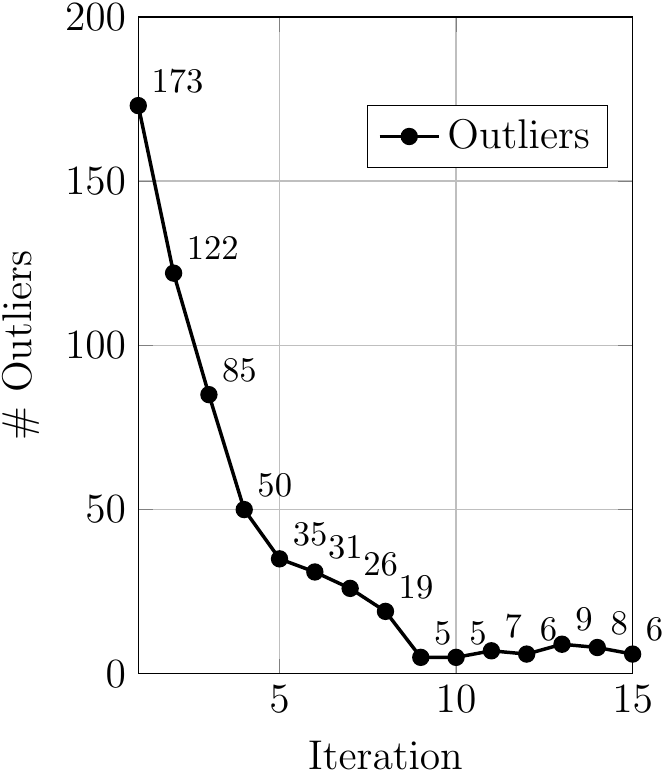}
\caption{Left: imbalance at the $9$th iteration of the balancing process. Right: number of outliers along the balancing process.}
\label{fig:out}	
\end{figure*}

 In  Figure~\ref{fig:conv}, we observe a level of imbalance (around $4\%$) from which the convergence worsens,
 although eventually it reduces down to $0.8\%$.
 In Figure~\ref{fig:out} (left) we show the normalized time per rank at the $9th$ iteration of the balancing process. 
 Note that most of the ranks present a value between $0.99$ and $1.01$, but five ``outliers'' remain. 
 The evolution of the number of outliers (processes with more than $1\%$ deviation with respect to the average) along the balancing 
 process appears in the right part of the figure. 
 We observe that it decreases exponentially during the first $9$ iterations, but then it oscillates between $5$ and $9$. 
 With further analysis of the results, we have observed that these outliers seem to appear randomly, which is probably caused by 
 system performance variability.

\begin{figure}[!h]
  \centering
    \includegraphics[width=0.4\textwidth]{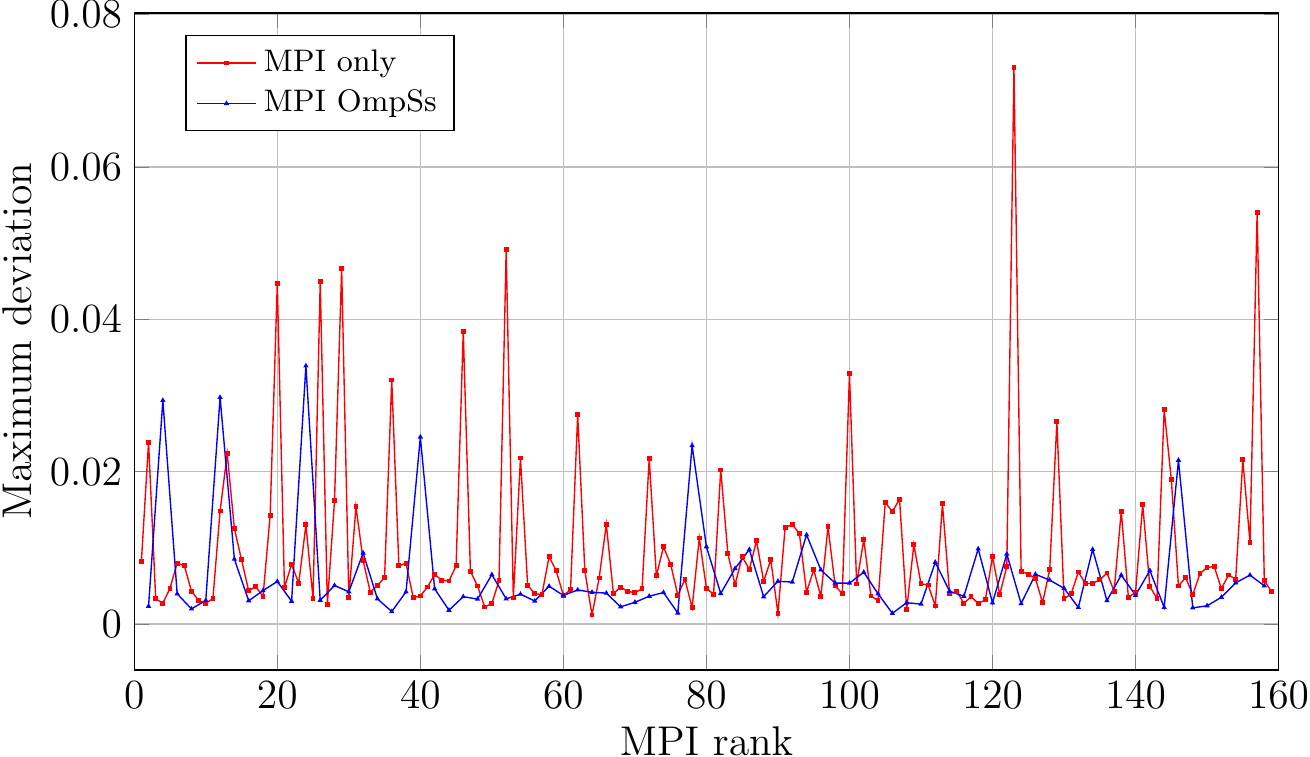}
\caption{Maximum time deviation for executions with the same configuration (31.5M elements mesh). Comparison of MPI vs MPI+OmpSs.}
\label{fig:var}	
\end{figure}

We have tested the system performance variability for a mesh of $31.5$M covering the same JAXA geomtry executed in $4$ nodes. 
The afore-used MPI+OmpSs configuration has been compared with the pure MPI option. 
We have launched each execution $10$ times using the same nodes and partition. In a perfectly stable system the time per execution 
and per MPI rank would be constant. Figure~\ref{fig:var} shows the maximum deviation per MPI-rank with respect to its average. We 
observe that it reaches 3.4$\%$ for the MPI+OmpSs configuration and up to 7$\%$ for the pure MPI one. The difference between both 
configuration is something expected. A possible slowdown in a CPU-core can be compensated in the shared memory space when using OmpSs.
Indeed it is divided exactly by two, that is the number of OmpSs threads per MPI process. The instability of the system is our main 
reason to chose the MPI+OmpSs configuration for the CPU executions.

\begin{figure*}[!h]
  \centering
    \includegraphics[width=0.6\textwidth]{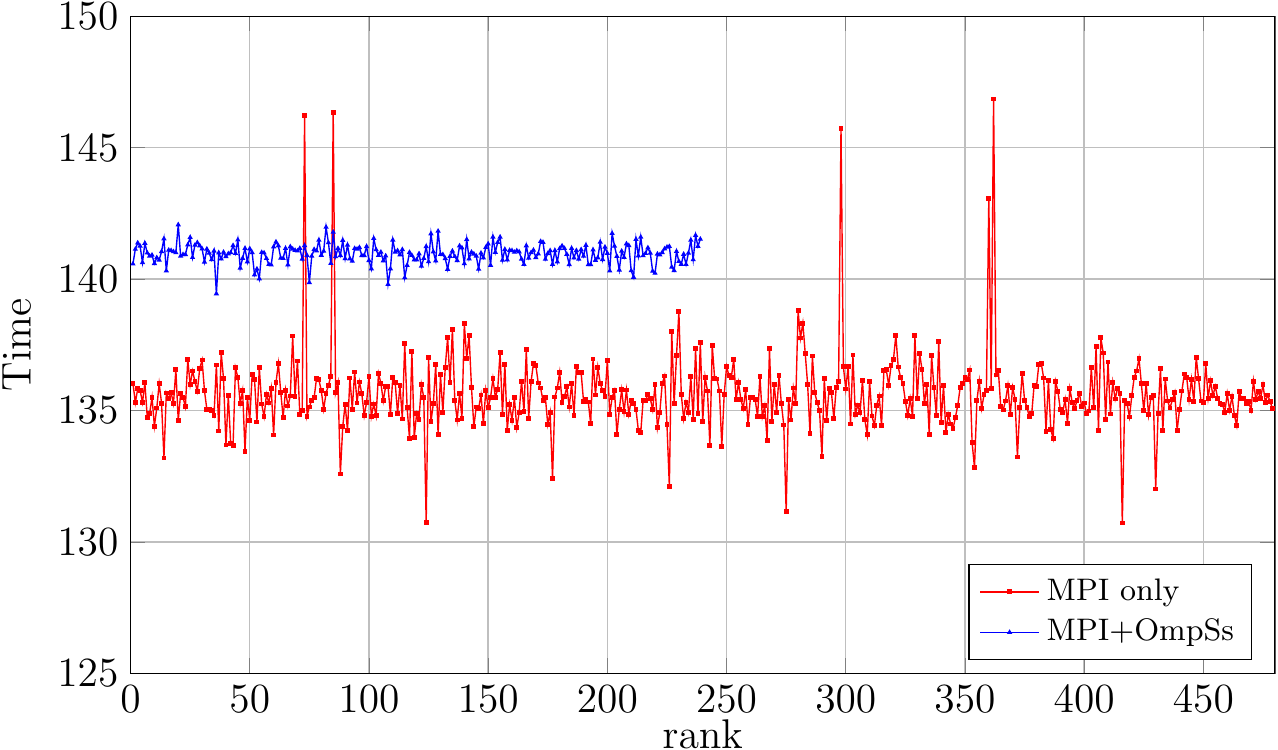}
\caption{Left: imbalance at the $15$th iteration of the balancing process.}
\label{fig:ups}	
\end{figure*}

In Figure~\ref{fig:ups} we show the normalized time per rank at the $15th$ iteration of the balancing process for the  
176M elements mesh. We compare again the pure-MPI versus the MPI+OmpSs execution. The number of CPU cores engaged is the same  
in both cases, but the number of subdomains of the MPI+OmpSs configuration halves the othe,  because two thread
are launched per  MPI-process. We observe that the average time  across the MPI-processes
is lower for the pure-MPI execution but, as this configuration is less resilient to the system noise, the execution time - which is 
dictated by the maximum time - is higher.

Note that our balancing  algorithm adjusts the partition dynamically  but it is not a runtime mechanism. That is,
we can not expect to correct the random system noise through the partition correction. This could be achieved by means of complementary runtime 
mechanisms. 
The robustness of the algorithm lies in the fact that it converges to a structurally balanced partition, a runtime mechanism can 
be very efficient upon. As aforementioned, this 
robustness comes from the fact that in the regression used to define the correction coefficients all previous measurments are taken into account.

\section{Concluding Remarks}
\label{sec:conclusions}

This paper presents an improvement of our in-house parallel mesh partitioner based on the Hilbert Space-Filling Curve. 
The former algorithm was highly efficient, but the number of parallel processes usable was restricted to powers of 4  and 8 
for the 2D and 3D case, respectively. We have overcome such restriction, and we have shown some preliminary tests performed 
on the MareNostrum IV supercomputer. Moreover, we have demonstrated the applicability of the partition algorithm for dynamic 
load-balancing. The strategy consists of monitoring the imbalance at runtime and adjusting the subdomains
accordingly. The new target weights are evaluated through a linear regression taking into account all the previous measurements; 
this approach makes the balancing process resilient to outlier measurements. We have demonstrated the accuracy of the balancing approach 
for a 176M elements hybrid mesh around the JAXA Standard Model (JSM) for airplane simulations.  On a pure-CPU execution, where the
imbalance is caused by the mesh heterogeneity, the final balance achived using 12 nodes of the \power{} installed at BSC is  
above $99\%$.

 \section*{Acknowledgment}
This work is partially supported by the BSC-IBM Deep Learning Research Agreement, under JSA “Application porting, analysis 
and optimization for POWER and POWER AI”.
It has also been partially supported by the EXCELLERAT project funded by the European Commission’s ICT activity of 
the H2020 Programme under grant agreement number: 823691.
It has also received funding from the European Union's Horizon 2020 research and innovation programme under grant agreement 
number: 846139 (Exa-FireFlows).
This paper expresses the opinions of the authors and not necessarily those of the European Commission. 
The European Commission is not liable for any use that may be made of the information contained in this paper.
This work has also been financially supported by the Ministerio de Economia, Industria y Competitividad, of Spain (TRA2017-88508-R).
The computing experiments of this paper have been performed on the resources of the Barcelona Supercomputing Center. .

\end{document}